\documentclass{aa}

\usepackage{url}
\usepackage{natbib}
\usepackage{array}
\usepackage[bf]{subfigure}
\usepackage[varg]{txfonts}
\usepackage{graphicx}

\newcommand{\beq}{\begin{equation}}
\newcommand{\eeq}{\end{equation}}
\newcommand{\bdi}{\begin{displaymath}}
\newcommand{\edi}{\end{displaymath}}

\newcommand{\herschel}{\textit{Herschel}}
\newcommand{\getsources}{\textsl{getsources}}
\newcommand{\getfilaments}{\textsl{getfilaments}}

\newcommand{\micron}{$\mu$m}
\newcommand{\nh}{$N_{\mathrm{H}_2}$}

\newcommand{\mline}{\textit{M}$_{\mathrm{line}}$}

\newcommand{\mtot}{\textit{M}$_{\mathrm{line,tot}}$}
\newcommand{\mcore}{\textit{M}$_{\mathrm{line,core}}$}
\newcommand{\mwing}{\textit{M}$_{\mathrm{line,wing}}$}

\newcommand{\mcrit}{\textit{M}$_{\mathrm{crit}}$}

\newcommand{\mpc}{\textit{M}$_{\mathrm{\odot}}$\,pc$^{-1}$}
\newcommand{\av}{$A_{\mathrm{V}}$}

\newcommand{\fsample}{core-scale}
\newcommand{\core}{core}

\begin{document}

%\shorttitle{Galactic Cold Cores VII}
%\shortauthors{Rivera-Ingraham, A.~et al.}

\title{Galactic Cold Cores\thanks{\MakeLowercase{\textit{\MakeUppercase{h}erschel} is an \MakeUppercase{ESA} space observatory with science instruments provided by \MakeUppercase{E}uropean-led \MakeUppercase{P}rincipal \MakeUppercase{I}nvestigator consortia and with important participation from \MakeUppercase{NASA}.}} VIII: Filament Formation and Evolution}
\subtitle{Filament Properties in Context with Evolutionary Models}

%\subtitle{Galactic Cold Cores VII} 

\author{A.~Rivera-Ingraham\inst{1}\and 
I.~Ristorcelli\inst{2,3}\and
M.~Juvela\inst{4}\and
J.~Montillaud\inst{5}\and
A.~Men'shchikov\inst{6}\and
J.~Malinen\inst{4}\and 
V.-M.~Pelkonen\inst{5,4,7}\and
A.~Marston\inst{1}\and
P.~G.~Martin\inst{8}\and
L.~Pagani\inst{9,10}\and
R.~Paladini\inst{11}\and
D.~Paradis\inst{2,3}\and
N.~Ysard\inst{12}\and
D.~Ward-Thompson\inst{13}
J.-P.~Bernard\inst{2,3}\and
D.~J.~Marshall\inst{6}\and
L.~Montier\inst{2,3}\and
L.~V.~T\'oth\inst{14}
}

\offprints{A. Rivera-Ingraham, \email{alana.rivera@esa.int}}

\institute{European Space Astronomy Centre (ESA/ESAC), Operations Department, Villanueva de la Ca\~nada (Madrid), Spain
\and
Universit\'{e} de Toulouse; UPS-OMP; IRAP;  Toulouse, France
\and
CNRS; IRAP; 9 Av. colonel Roche, BP 44346, F-31028 Toulouse cedex 4, France
\and
Department of Physics, P.O. Box 64, FI-00014, University of Helsinki, Finland
\and
Institut UTINAM, CNRS 6213, OSU THETA, Universit\'{e} de Franche-Comt\'{e},41 bis avenue de l'Observatoire, 25000 Besan\c{c}on, France
\and
Laboratoire AIM, CEA/DSM/Irfu –- CNRS/INSU –- Universit\'e Paris Diderot, CEA-Saclay, F-91191 Gif-sur-Yvette Cedex, France
\and
Finnish Centre for Astronomy with ESO, University of Turku, V\"ais\"al\"antie 20, 21500 Piikki\"o, Finland
\and
Canadian Institute for Theoretical Astrophysics, University of Toronto, 60 St. George Street, Toronto, ON M5S~3H8, Canada
\and
LERMA, Observatoire de Paris, PSL Research University, CNRS, UMR 8112, F-75014 Paris, France
\and
Sorbonne Universit\'es, UPMC Univ. Paris 6, UMR 8112, LERMA, F-75005, Paris, France
\and
Infrared Processing Analysis Center, California Institute of Technology, 770 S. Wilson Ave., Pasadena, CA 91125, USA
\and
IAS, CNRS (UMR8617), Universit\'{e} Paris Sud, Bat. 121, F-91400 Orsay, France
\and
Jeremiah Horrocks Institute, University of Central Lancashire, Preston, Lancashire, PR1 2HE, UK
\and
E\"otv\"os University, Department of Astronomy,P\'azm\'any P. s. 1/a,H-1117, Budapest, Hungary
}

\date{Received X September 2016 / Accepted X September 2016}

%%%%%%%%%%%%%%%%%%%%%%%%%%%%%%%%%%%%%%%%%%%%%%%%%%%%%%%%%%%%%%%%%%%%%%%%%%%%

\abstract 
%CONTEXT
{The onset of star formation is intimately linked with the presence of massive, unstable filaments. These structures are therefore key for theoretical models aiming to reproduce the observed characteristics of the star formation process.}
%AIMS
{As part of the filament study carried out by the \herschel\ Galactic Cold Cores Key Programme, here we study the filament properties presented in GCC VII (Paper I) in context with theoretical models of filament formation and evolution.}
%METHODS
{A conservative sample of filaments at a distance $D<500$\,pc was extracted with the \getfilaments\, algorithm. The physical structure of the filaments was quantified according to two main components: the central (Gaussian) region (\core\ component), and the power-law like region dominating the filament column density profile at larger radii (wing component). The properties and behaviour of these components relative to the total linear mass density of the filament and its environmental column density were compared with the predictions from theoretical models describing the evolution of filaments under gravity-dominated conditions.}
%RESULTS
{The feasibility of a transition from a subcritical to supercritical state by accretion is dependent on the combined effect of filament intrinsic properties and environmental conditions. Reasonably self-gravitating (high \mcore) filaments in dense environments (\av$\approx$3\,mag) can become supercritical in timescales of $t\sim1$\,Myr by accreting mass at constant or decreasing width. The trend of increasing \mtot\ (\mcore\ and \mwing), and ridge \av\ with background also indicates that the precursors of star-forming filaments evolve coevally with their environment. The simultaneous increase of environment and filament \av\ explains the association between dense environments and high \mcore\ values, and argues against filaments remaining in constant single-pressure equilibrium states. The simultaneous growth of filament and background in locations with efficient mass assembly, predicted in numerical models of collapsing clouds, presents a suitable scenario for the fulfillment of the combined filament mass$-$environment criterium that is in quantitative agreement with \herschel\ observations.}
{}

\keywords{ISM: clouds --- infrared: ISM --- submillimeter: ISM --- dust,extinction --- Stars:formation}
\maketitle

%%%%%%%%%%%%%%%%%%%%%%%%%%%%%%%%%%%%%%%%%%%%%%%%%%%%%%%%%%%%%%%%%%%%%%%%%%%%

\section{Introduction} \label{sec:intro}
Filamentary structures are a wide-spread phenomenon in the interstellar medium (ISM). 
They constitute a complex hierarchical population with a wide range of properties in terms of 
crest (ridge) column density, linear mass density (\mline), and length (e.g.,  \citealp{hacar2013}; \citealp{hennemann2012}).
Observations carried out with the \herschel\ Space Observatory (\herschel; \citealp{pilbratt2010}) 
 indicate that filaments in the local neighbourhood might be characterised by 
a quasi-constant average width of $\sim$$0.1$\,pc  (e.g., \citealp{arzoumanian2011}), although 
with a wide spread around this value and possibly even increasing at larger distances (e.g., \citealp{juvela2012a}; \citealp{schisano2014}). 
Filaments are also ubiquitous in a wide 
range of environments and conditions, from the most diffuse regions at high galactic latitudes 
(e.g.,  \citealp{miville2010}) to the active high-mass star forming complexes (e.g., \citealp{hennemann2012}).

Despite their diversity, the observed tendency of prestellar cores and young stellar objects (YSOs) to be 
associated with the densest and most massive of these filaments (supercritical: \mline$ > $\mcrit, 
where \mcrit$=2c_{\mathrm{s}}^2/G$$\sim$16.5\,\mpc\ for a dust temperature of $T\approx10$\,K; 
e.g., \citealp{inutsuka1992}; \citealp{andre2010}) has made them a critical component that needs to be accurately constrained
for the development of viable observational and theoretical models of star formation. 
Processes such as global cloud collapse (e.g., \citealp{peretto2013}), convergence of 
large scale flows (e.g., \citealp{schneider2010b}), or cloud-cloud collisions (e.g., \citealp{duarte2011}) are 
just examples of the various scenarios invoked to account for the presence of filaments in star-forming regions.

In \citet{rivera2016} (Paper I hereafter), the data acquired by the \herschel\ Galactic Cold Cores Key Programme 
(GCC; P.I: M. Juvela; \citealp{juvela2012a}) was used to extract a sample of filaments for 
Galactic fields located at $D\le500$\,pc. This study complemented the work presented 
in  \citet{juvela2012a}, and aimed to provide a robust characterisation of the physical properties 
of the filament population under different environmental conditions. 
The goal of this work is to apply and interpret the results derived in Paper I in context 
with the predictions from gravity-dominated models of star formation. 
Comparison between the observed filament properties and 
predictions from such models will lead to a better understanding of the processes and 
conditions needed for the formation of star-forming filaments in (gravity-driven) scenarios.
The role of external events in filament formation such as external feedback and collisions are 
the subject of ongoing work and will be published in an upcoming study.

This work is organised according to the following structure:
In Sect. \ref{sec:data} and Sect. \ref{sec:methods} we provide a brief summary of the 
datasets and methods used in Paper I to identify and characterise the filament sample. 
The key properties extracted from the filament catalog are included in Sect. \ref{sec:results}. 
These results are analysed and discussed in Sect. \ref{sec:discussion} in order to constrain 
the formation process of star-forming supercritical according to accretion-based models.
Our conclusions are listed in Sect. \ref{sec:conclusion}.

%%%%%%%%%%%%%%%%%%%%%%%%%%%%%%%%%%%%%%%%%%%%%%%%%%%%%%%%%%%%%%%%%%%%%%%%%%%%%%%%%%%%%

\section{\herschel\ Maps \& Data Processing} \label{sec:data}
The \herschel\ maps with filament detections and used as a base in Paper I constitute a subsample of 38 fields out of the 116 regions observed 
by the GCC Programme (see \citealp{juvela2012a} and \citealp{montillaud2015} for a detailed description of the processing and map 
properties). 

The SPIRE ($250$\,\micron, $350$\,\micron, and $500$\,\micron; \citealp{griffin2010}) 
and PACS $160$\,\micron\ maps \citep{pog2010} were processed with the Herschel Interactive Processing 
Environment (HIPE\footnote{HIPE is a joint 
development by the Herschel Science Ground Segment Consortium, 
consisting of ESA, the NASA Herschel Science Center, and 
the HIFI, PACS and SPIRE consortia.}) v.10.0 and the \textsl{Scanamorphos} package 
version 20 \citep{roussel2013}, respectively.

Column density and temperature maps were produced from the colour and offset 
corrected SPIRE brightness maps convolved to a common resolution of $40$\arcsec.
The pixel-by-pixel fitting of spectral energy distributions 
(SEDs) was carried out assuming a dust 
opacity of $0.1$\,cm$^2$\,g$^{-1}$ at $1$\,THz \citep{hildebrand1983} for a 
fixed dust emissivity index of $\beta=2$. We assumed a mean atomic weight per molecule of $\mu=2.33$ 
for consistency with previous filament studies.

%%%%%%%%%%%%%%%%%%%%%%%%%%%%%%%%%%%%%%%%%%%%%
\section{Methodology: Detection \& Characterisation of the Filament Sample} \label{sec:methods}
The target structures of our analysis consisted of the most robust filamentary detections located at $D<500$\,pc and 
directly linked to the formation of cores within the resolution limitations of the data 
(full with at half maximum: FWHM$_{\mathrm{core}}<0.2$\,pc). 
The process of filament extraction, selection, and analysis, are outlined below, and we refer to Paper I for a 
detail description of the process and techniques employed.

A preliminary catalog of filament detections in each field was obtained from the \nh\ maps using the \getfilaments\, algorithm v1.140127 \citep{getfilaments}. 
This initial extraction was carried out as part of the source detection process performed with the 
multi- scale, multi-wavelength source extraction algorithm \getsources\ \citep{getsources2012}. 
During the filament detection process, \getfilaments\ effectively identified and separated all the filamentary contribution in the map from that associated with 
compact sources and background/noise fluctuations, providing a set of images with 'clean' filament profiles from which 
physical parameters such as width, length, and intensity can be derived.
The algorithm also quantified, for each filament pixel, the fraction of the total intensity associated with different spatial scales in the image.
Only those filaments with significant emission in the spatial scales relevant for the formation of prestellar cores ($<0.2$\,pc at $D\le500$\,pc) 
were considered to be `core-bearing' filaments ( `\fsample' filaments) and selected for further analysis.

Characterisation of the filament population was carried out by examining the structural and environmental properties of each filament.
The average radial column density profile of each detection (including contribution from compact sources to the profile) 
was fitted with an idealized model of a Plummer-like (\citealp{whitworth2001}; \citealp{nutter2008}) 
cylindrical filament using  a non-linear least squares minimization routine. Integration of the best-fit profile yielded an estimate of the total 
linear mass density (\mtot). Following the approach used in previous studies, the filament profile was further assumed to 
be characterised by a Gaussian-like inner region comprising the most central and densest parts of the filament. This component was 
described in Paper I as the filament `\core-component', and defines the characteristic FWHM of the filament. 
The  `wing-component' is dominated by the power-law like region of the filament profile at large radii, so that the total 
linear mass density can be expressed as \mtot$=$\mcore$+$\mwing. 
The mean intrinsic column density of the filament was estimated by averaging the values of the pixels associated with the filament crest. 
An estimate of the background level was obtained by averaging the same pixels, but with \nh\ values measured from the 
complementary background map obtained by \getfilaments\ (\getsources) during the filament extraction process. 

\begin{table*}[t!]
\caption{Filament parameters of the regimes for the \fsample\ filament sample.}
\label{table:regimes}
\centering
\begin{tabular}{l l l l l l l l}
\hline \hline
Regime&$<$\mcore$>$&$<$\mwing$>$&$<$\nh$>^a$&$<$\av$>^{a,b}$&$<$BKG \nh$>$&$<$BKG \av$>^b$&FWHM\\ 
&[\mpc]&[\mpc]&[$10^{20}$\,cm$^{-2}$]&[mag]&[$10^{20}$\,cm$^{-2}$]&[mag]&[pc]\\
\hline
1 ALL&$2.16\pm0.22$&$2.40\pm0.70$&$8.92\pm0.82$&$0.95\pm0.09$&$13.77\pm2.05$&$1.46\pm0.22$&$0.11\pm0.01$\\
1 ALL(HB)$^c$&$2.43\pm0.40$&$3.29\pm1.98$&$12.34\pm2.02$&$1.31\pm0.21$&$26.90\pm2.53$&$2.86\pm0.27$&$0.09\pm0.01$\\
1 ALL(LB)$^c$&$2.08\pm0.26$&$2.13\pm0.72$&$7.86\pm0.69$&$0.84\pm0.07$&$9.73\pm1.01$&$1.04\pm0.11$&$0.11\pm0.01$\\
\hline
2 ALL&$5.97\pm0.25$&$7.72\pm1.38$&$23.02\pm3.38$&$2.45\pm0.36$&$25.60\pm2.57$&$2.72\pm0.27$&$0.14\pm0.01$\\
2 ALL(HB)&$5.97\pm0.27$&$6.47\pm2.04$&$28.51\pm5.80$&$3.03\pm0.62$&$35.35\pm1.48$&$3.76\pm0.16$&$0.11\pm0.02$\\
2 ALL(LB)&$5.96\pm0.44$&$9.11\pm1.84$&$16.93\pm1.85$&$1.80\pm0.20$&$14.77\pm0.92$&$1.57\pm0.10$&$0.18\pm0.02$\\
2 SB$^d$&$6.10\pm0.30$&$3.57\pm0.73$&$21.04\pm2.54$&$2.24\pm0.27$&$25.36\pm3.27$&$2.70\pm0.35$&$0.15\pm0.02$\\
2 SB (HB)&$5.77\pm0.30$&$2.69\pm0.83$&$22.80\pm3.87$&$2.43\pm0.41$&$33.91\pm1.89$&$3.61\pm0.20$&$0.13\pm0.02$\\
2 SB (LB)&$6.58\pm0.56$&$4.81\pm1.17$&$18.58\pm2.95$&$1.98\pm0.31$&$13.40\pm1.29$&$1.43\pm0.14$&$0.18\pm0.03$\\
2 SP$^d$&$5.73\pm0.44$&$14.83\pm0.74$&$26.43\pm8.35$&$2.81\pm0.89$&$26.01\pm4.51$&$2.77\pm0.48$&$0.14\pm0.03$\\
2 SP (HB)&$6.45\pm0.56$&$15.29\pm1.47$&$41.85\pm16.54$&$4.45\pm1.76$&$38.71\pm0.32$&$4.12\pm0.03$&$0.08\pm0.03$\\
2 SP (LB)&$5.20\pm0.55$&$14.49\pm0.85$&$14.87\pm1.86$&$1.58\pm0.20$&$16.48\pm0.76$&$1.75\pm0.08$&$0.18\pm0.03$\\
\hline
3 ALL&$13.61\pm1.42$&$20.26\pm5.19$&$52.18\pm8.12$&$5.55\pm0.86$&$27.70\pm4.27$&$2.95\pm0.45$&$0.12\pm0.01$\\
3 ALL(HB)&$15.14\pm1.57$&$23.64\pm7.28$&$58.16\pm11.03$&$6.19\pm1.17$&$32.56\pm4.65$&$3.46\pm0.49$&$0.13\pm0.02$\\
3 ALL(LB)&$10.54\pm1.18$&$13.51\pm4.25$&$40.21\pm6.67$&$4.28\pm0.71$&$17.99\pm1.64$&$1.91\pm0.17$&$0.12\pm0.00$\\
\hline
\multicolumn{8}{l}{{$^a$ Average intrinsic (background-free) \nh\ of crest and standard error on the mean.}}\\ 
\multicolumn{8}{l}{{$^b$ \nh$=9.4\times10^{20}$\,cm$^{-2}$ \av$/$mag.}}\\ 
\multicolumn{8}{l}{{$^c$ HB$=$high-background; LB$=$low-background.}}\\ 
\multicolumn{8}{l}{{$^d$ SB$=$subcritical; SP$=$supercritical.}}\\ 
\hline
\end{tabular}
\end{table*}

\begin{figure*}[ht]
\centering
\subfigure[]{%
\label{fig:model1}
\includegraphics[scale=0.40,angle=0]{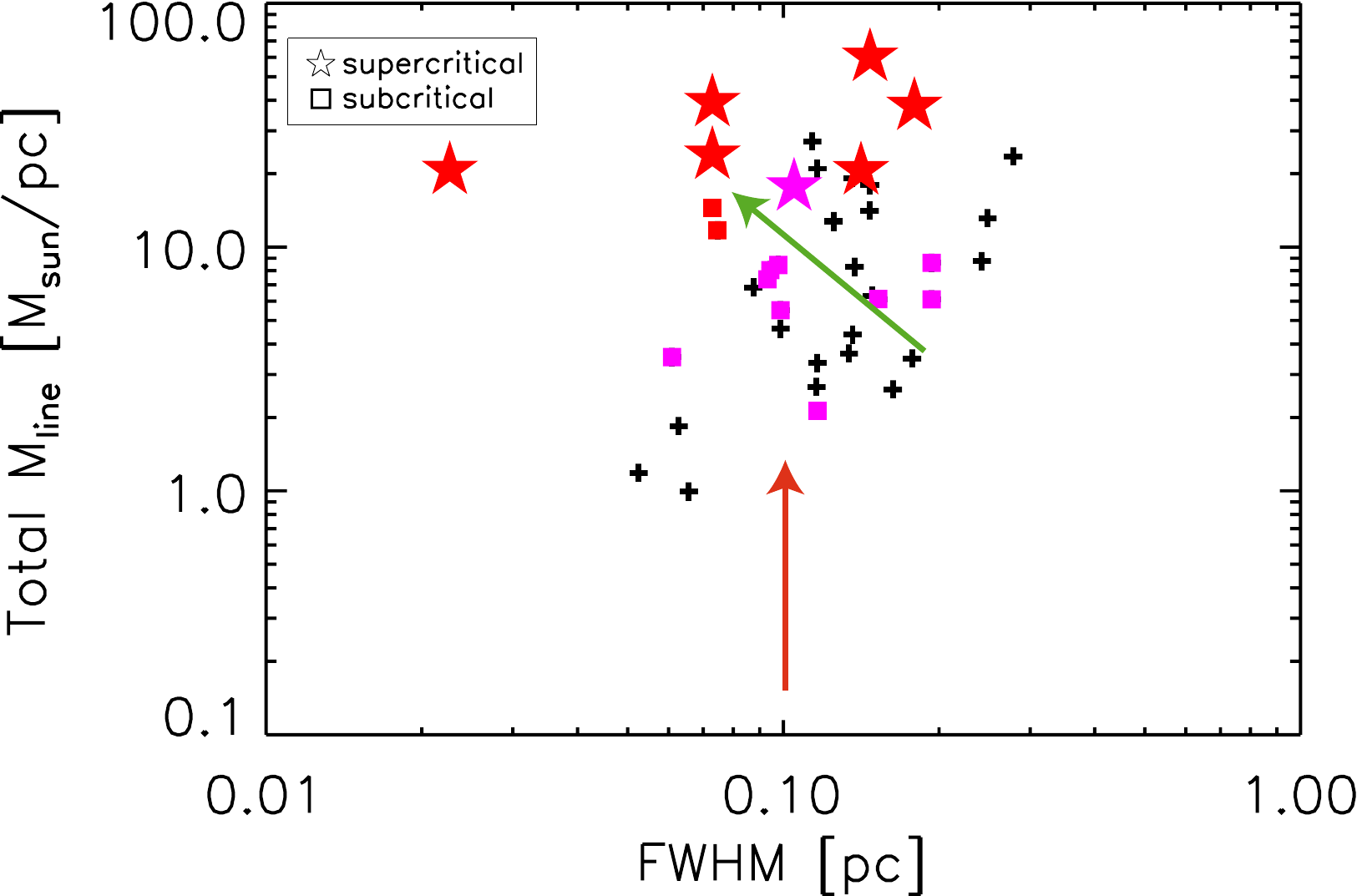}
}%
\subfigure[]{%
\label{fig:model2}
\includegraphics[scale=0.40,angle=0]{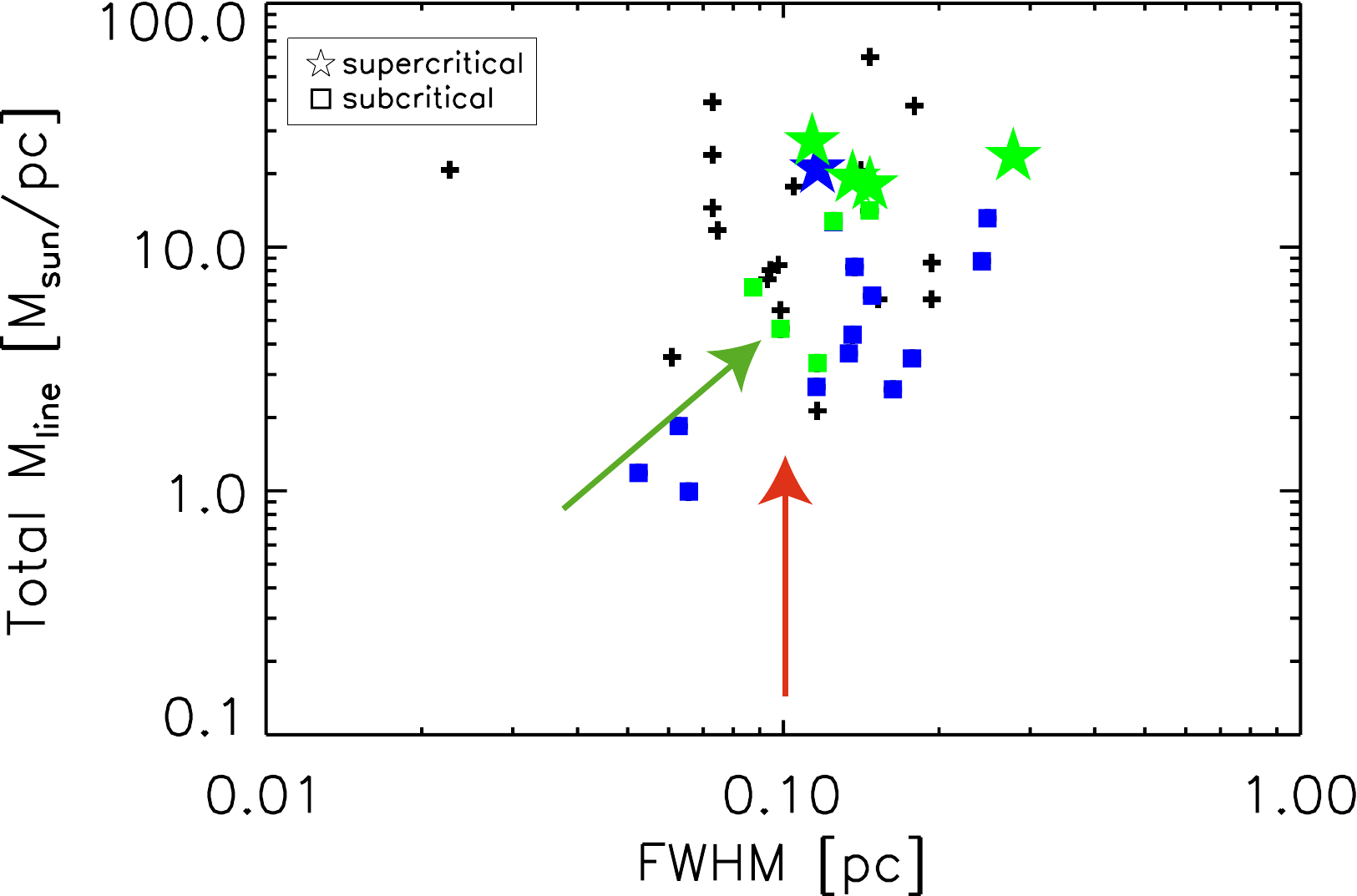}
}\\%
\subfigure[]{%
\label{fig:model3}
\includegraphics[scale=0.40,angle=0]{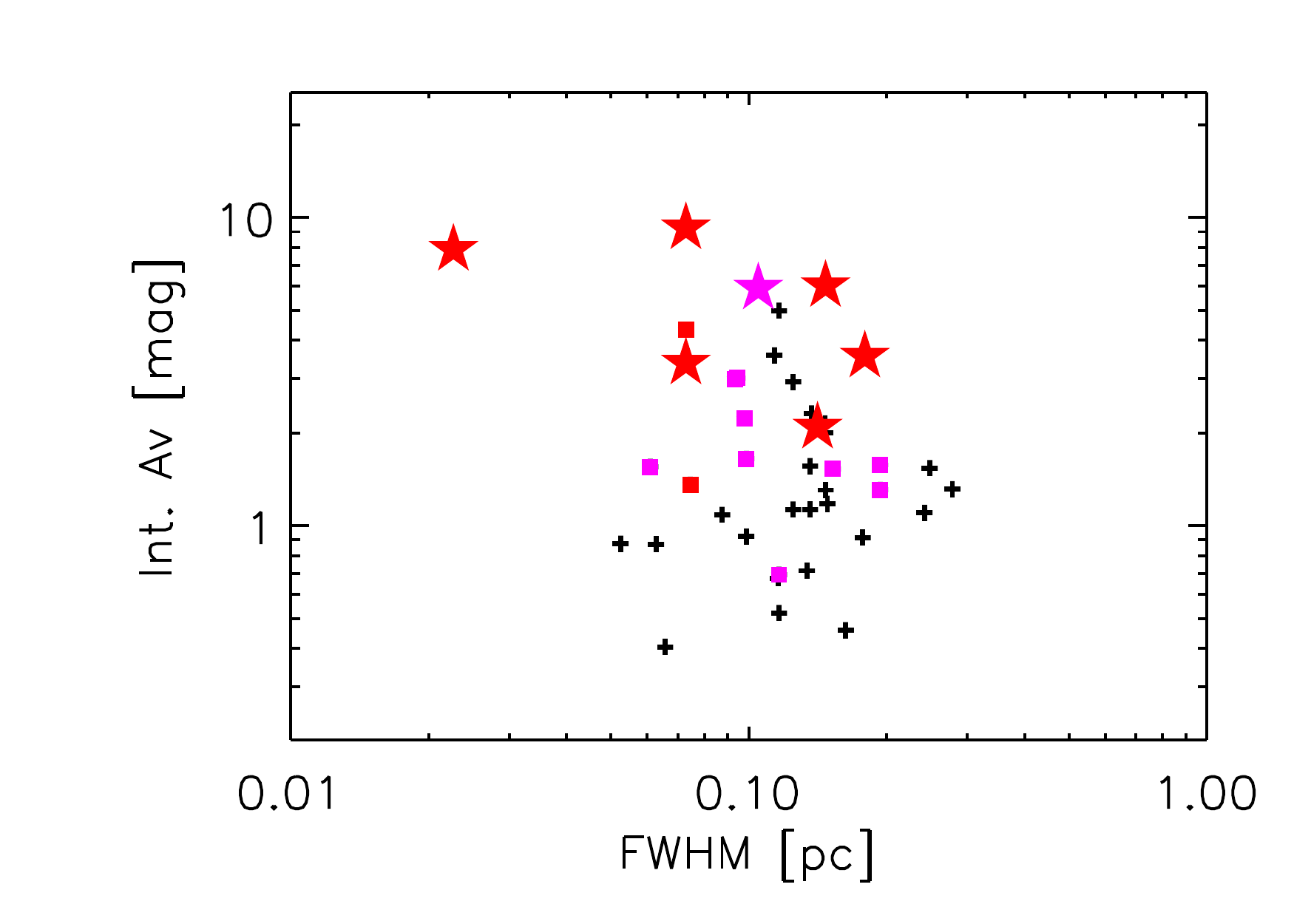}
}%
\subfigure[]{%
\label{fig:model4}
\includegraphics[scale=0.40,angle=0]{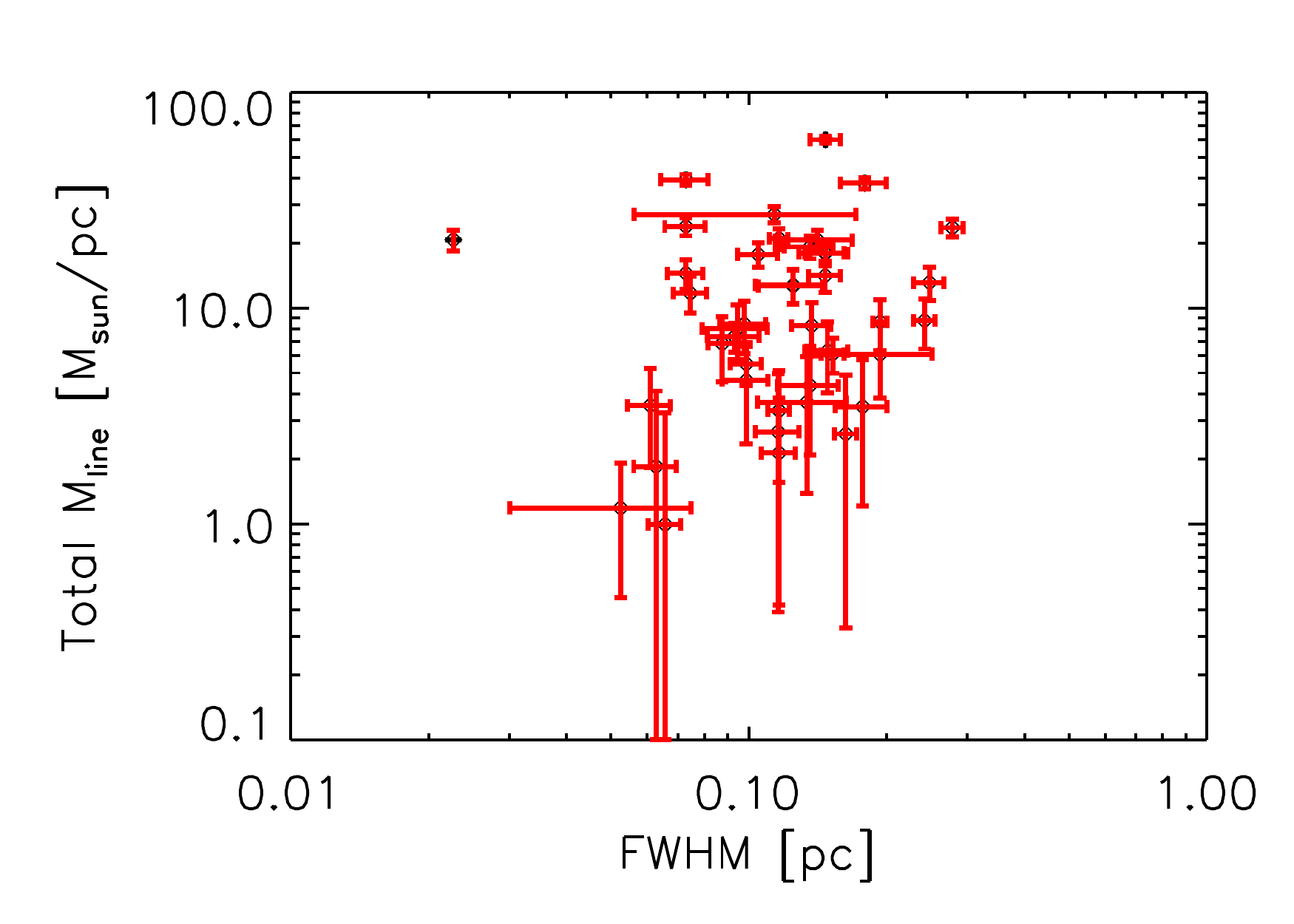}
}\\%
\caption{Filament characteristic intrinsic total linear mass density (\mtot) as a function of FWHM  for HBs (filled red/magenta symbols; [a]) and LBs (filled blue/green symbols [b]) of the \fsample\ filament sample. Panel (c) shows the \av-FWHM distribution for HBs. As reference, black crosses in Figs. \ref{fig:model1} and \ref{fig:model2} mark the position of filaments in low-backgrounds and high-backgrounds, respectively. Panel (d) is the same as (a) and (b), but with uncertainties included as reference. The red arrow indicates the evolutionary direction at constant $\sim0.1$\,pc width (e.g., Arzoumanian et al. 2011; 2013). The green arrow indicates an alternative evolutionary path leading to filaments with the same final properties and conveying similar changes as the original model, but allowing for a (conservative) change in filament width (see text). Separation according to the relative contribution of \core/wing components to the filament \mtot\ is marked as follows: Core-dominated (blue/magenta for LBs and HBs, respectively), and wing-dominated (green/red). } 
%}
\label{fig:evolution}
\end{figure*}

%%%%%%%%%%%%%%%%%%%%%%%%%%%%%%%%%%%%%%%%%
\section{Results: Potential Evolutionary Trends for Accretion-based Models}\label{sec:results}
A total of $14$ fields contained filaments in the local neighbourhood 
satisfying the intrinsic and reliability criteria established for the selection of \core-bearing filaments. 
These filaments were classified according to the component dominating their total linear 
mass density (core-dominated if \mcore/\mwing$>1$, wing-dominated otherwise),  
stability (supercritical if \mtot$\ge$\mcrit, subcritical otherwise), and environmental level 
(low-background (LB) filaments if background level below the mean of the population of 
\av$\approx2.2$\,mag, high-background (HB) if 
above this value\footnote{\nh$=9.4\times10^{20}$\,cm$^{-2}$ \av$/$mag; \citet{bohlin1978}}).
 We used these classifications to identify not only global properties of the filament population, but 
 also potential trends and correlations that could provide insight into the requirements, 
 limitations, and nature of the physical processes driving filament formation and evolution. 
  
Three major filament  `regimes' are identified (see Paper I): a core-dominated subcritical regime ( Regime 1; \mcore$<4.2$\,$M_{\odot}$\,pc$^{-1}$), 
a transition regime comprising a mixture of core-dominated, wing-dominated, 
subcritical, and supercritical filaments (Regime 2; $4.2\le$\mcore$\le8.4$\,$M_{\odot}$\,pc$^{-1}$), and a regime exclusively associated with 
supercritical filaments (Regime 3; \mcore$>8.4$\,$M_{\odot}$\,pc$^{-1}$). 
These regimes, initially selected according to their \mcore, show distinct average structural properties 
(\mtot, crest \nh) and environment. This can be clearly observed in the global mean properties of 
each regime highlighted in Paper I (Tables 3 and 4), and in more detail in Table \ref{table:regimes} of this work.  

Filament growth and mass assembly via accretion, as predicted by theoretical models
and observational studies (e.g., \citealp{schisano2014}), would be 
characterised by an increase in \mtot\ with time. 
Just based on the average global characteristics of the different regimes
we were able to conclude in Paper I that both components (\mcore\ and \mwing)  
appear to increase with increasing \mtot\ (i.e., with time), with the wing component 
dominating over the core component at high \mtot\ (late stages) of evolution. 
We were also able to infer  that the most 
massive filament components are primarily associated with the densest environments. 
Our results clearly indicated that local external conditions must play a fundamental role 
in filament evolution. 

Table \ref{table:regimes} provides a detailed list of the average properties of the different filament 
groups (classified according to stability and background \nh\ level) within 
each of the three regimes. 
The potential changes in filament structure as a function of time (\mtot) and environment 
quantified in this Table are visually illustrated in Fig. \ref{fig:evolution}.

In a transition from an overall subcritical to supercritical regime, 
filaments are characterised by an increase in central ridge column density 
(e.g., Fig. \ref{fig:model3}).
This would be a natural outcome for structures undergoing collapse and substructure 
growth via gravitational forces, especially in dense environments where such processes would be more significant 
due to the availability of material.
Our results also indicate a roughly constant or possibly moderate 
decrease in filament width with time for HB filaments (Fig. \ref{fig:model1}), and 
an increase for LB filaments (Fig. \ref{fig:model2}).  
These figures indicate that if there are indeed changes in width during filament evolution from a 
subcritical to supercritical regime, the small variations associated with such changes would  
explain the apparently constant value of this quantity for filaments 
in radically different environments and star-forming states (e.g., \citealp{andre2010}). 
It would also support accretion as a major process in filament formation and 
evolution, as the presence of filaments with characteristic widths of $\sim0.1$\,pc 
have been theorised to be the result, at least in part, of accretion effects 
(e.g., accretion-driven turbulence and ion-neutral friction; \citealp{hennebelle2013a}).

%%%%%%%%%%%%%%%%%%%%%%%%%%%%%%%%%%%%%%%%%%%%%%%%%%%%%%%%%%%%%%%%%%%%
\section{Discussion: Filament Evolution under Gravitationally-Dominated Conditions}\label{sec:discussion}
Accretion is necessary for filament growth and, therefore, a key factor in any 
evolutionary model of filaments in the ISM. Indeed, 
accretion is predicted to play a crucial, if not dominant, role in any 
scenario of filament evolution.
Theoretical modelling of gravitational infall onto molecular filaments 
carried out by \citet{heitsch2013a} indicates that accretion cannot be prevented 
even in the presence of magnetic fields or turbulence. 
With timescales shorter than those of ambipolar and 
turbulent diffusion \citep{heitsch2014} 
accretion is one of the most likely candidates for driving the prestellar core and star formation process.

Evidence of mass assembly by accretion from the local environment has 
been readily observed for filaments associated with low and high-mass 
star formation alike (e.g., \citealp{hennemann2012}; \citealp{palmeirim2013}).
In the evolutionary model proposed in \citet{arzoumanian2011, arzoumanian2013}, 
accreting and collapsing supercritical self-gravitating filaments would increase their linear mass density
while keeping a constant filament width ($\sim0.1$\,pc). 
An increase in linear mass density by accretion for filaments associated with the prestellar and protostellar 
stages of star formation was also suggested in the work by \citet{schisano2014} 
and has been observed in numerical simulations. 
Filaments are a general intrinsic feature of molecular clouds (\citealp{smith2016}), 
therefore influenced by the evolution of their large scale environment. 
Extensive periods of accretion are predicted to arise due to global cloud collapse (e.g., \citealp{gomez2014}; \citealp{smith2014}), 
and the transition to a supercritical regime is an expected outcome in those regions in which 
the converging collapse leads to enhanced gathering of mass (\citealp{heitsch2009}, \citealp{smith2016}).
According to simple linear mass density stability criteria, 
subcritical filaments are defined as gravitationally unbound and stable against collapse, 
therefore most likely prone to dispersion before undergoing significant star formation. 
It is, however, still not observationally constrained how and under what conditions subcritical filaments evolve 
by accretion into an unstable supercritical regime, and how these observations fit in the proposed theoretical scenarios.

Based on their linear mass density and central crest column density, the GCC filaments 
comprise a key population in the intermediate region between 
the subcritical and supercritical states. 
These filaments also provide a statistically significant sample in a range of 
environments (e.g., Table \ref{table:regimes}). 
Our observations  can therefore be used to constrain a possible transition process between the two states, as well as 
the role of gravity and local environment in the formation, evolution, and ultimate fate of filaments  in the ISM.

\subsection{Assumptions for a Subcritical-Supercritical Transition Process via Accretion}
Several scenarios have been invoked that can explain the ubiquitous 
presence of filaments in the ISM. These are, for instance, the resulting structures 
caused by large-scale gravitational collapse of clouds formed by 
colliding flows or streams in the ISM (e.g., \citealp{burkert2004}), 
with initial density substructure determined  by turbulence and by thermal and dynamical instabilities 
(e.g., \citealp{heitsch2008a}), and with growth aided by dynamical as well as 
gravitational focussing. Indeed, filaments are predicted to appear as a natural consequence 
of the cloud formation process (e.g., \citealp{heitsch2009}).
Based on these and similar models, filament formation and evolution 
constitute a continuous process of which our \herschel\ observations are 
nothing more than snapshots in time. 
By definition, our observational sample comprises a set of structures with a well developed 
filamentary morphology. In consequence, each of the three filament Regimes 
(Table \ref{table:regimes})
is expected to contain filaments in different states of evolution. 
The use of of this sample to constrain filament evolution is based on the following key results and assumptions:

\begin{enumerate}

\item \textit{The filament population is dominated by structures with 
column densities high enough for gravity to dominate} (based on \nh\ estimates from e.g., \citealp{hartmann2001}).

The \herschel\ continuum datasets can therefore be used to broadly separate the 
filament sample according to their early or late evolutionary state, assuming that 
the general trends observed with increasing \mtot\ for the various filament parameters (Paper I) 
are driven by accretion and inflow of material. 

\item \textit{A supercritical filament in one Regime evolves from a subcritical filament 
by increasing its \mtot\ at the same time as its \mcore, \mwing, and ridge \nh}. 
These trends were observed and described in Paper I.

Numerical simulations (e.g., \citealp{burkert2004}; \citealp{hartmann2007}; \citealp{heitsch2008c}; \citealp{vazquez2009}; \citealp{smith2014}; \citealp{gomez2014}) as 
well as observational results (e.g., \citealp{schneider2010b}) support the long-range effects of gravity and therefore the 
potentially extensive reservoir of material that can, in principle, be available for accretion. 
For mean infall velocities of $v\sim0.8$\,km\,s$^{-1}$ (e.g., \citealp{heitsch2009}) 
accretion of material during the predicted timescale for prestellar core formation and collapse 
can in principle occur from regions located at $\ga1$\,pc from the filaments. 
The possibility of the environment evolving as the filament accretes implies that the precursor of a 
star-forming filament could be associated not only with a less massive structure, but 
also with a less dense background. This scenario is accounted for in our analysis by allowing 
filaments to evolve from a similar or more diffuse environment than that of the target supercritical structure.

\item \textit{The wing filament component dominates at a late stage of evolution}.  

Our results from Paper I indicate that 
in a transition from subcritical to supercritical state a filament increases its \mcore\ and \mwing. 
This simultaneous growth could be compatible with that predicted for other type of structures within 
clouds (e.g., cores; \citealp{naranjo2015}). 
Filaments with the highest \mtot\ have the most massive components \textit{and} are also predominantly wing-dominated. 
A potential precursor of a star-forming filament is therefore required to be core-dominated and with both \mcore\ and \mwing\ 
lower than that of the target wing-dominated supercritical filament. 
While the actual behaviour of the filament components with time requires an extensive analysis and modelling of 
filament properties under a range of conditions, our choice is justified based on the general trends of the filaments population observed in Paper I and the individual 
properties of each Regime (e.g., Table \ref{table:regimes}).  
In Regime 2, all supercritical filaments are wing-dominated and $\sim85$\% of the subcritical population is core-dominated. 
Only 2 out of the 6 supercritical filaments in Regime 3 are core-dominated.

\item \textit{Mechanisms capable of providing internal support, such as turbulence or magnetic fields, do not dominate filament evolution of the global filament population.}

Turbulence, for instance, is predicted to be highly linked to accretion and collapse (e.g., \citealp{burkert2004}), 
but these processes should still not be able to prevent accretion. 
Numerical models clearly indicate that  molecular clouds are dynamically evolving and contracting gravitationally despite the presence of magnetic fields \citep{ibanez2016}. The same process of large scale global collapse drives the evolution of the internal substructure of the cloud (filament, clump), therefore gravity and mass inflow become the principal and dominant mechanisms determining changes in filamentary properties as a function of time. Observational evidence supporting the driving role of gravity is found in the shape of the observed column density probability density functions of star-forming clouds, characterised by a power-law tail at the column densities typically associated with filamentary material and star-formation (e.g., \citealp{ballesteros2011}).

\end{enumerate}

By means of the above assumptions it is possible to quantify 
the changes associated with a transition from subcritical to supercritical state, 
if the filament was required to form new stars by accretion 
under the conditions predicted by observational and theoretical studies. 

\subsection{Formation and Evolution of Star-Forming Filaments}
\subsubsection{The "preferred" conditions of supercritical filaments}

\begin{figure}[ht]
\centering
\subfigure[]{%
\label{fig:theory1}
\includegraphics[scale=0.50,angle=0,trim=10cm 9cm 8cm 6cm, clip=true]{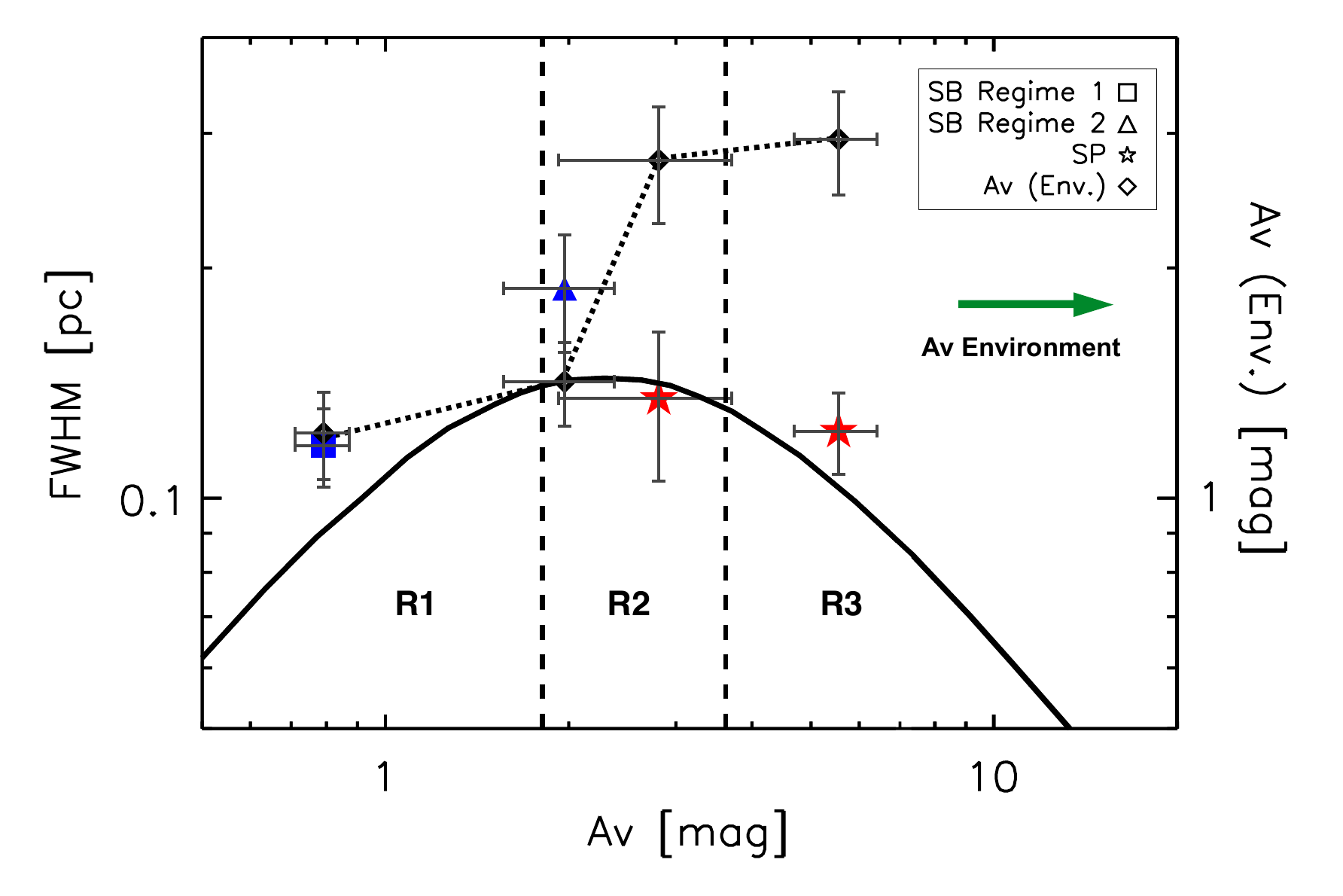}
}\\%
\subfigure[]{%
\label{fig:theory2}
\includegraphics[scale=0.50,angle=0,trim=1cm 0cm 0cm 0cm, clip=true]{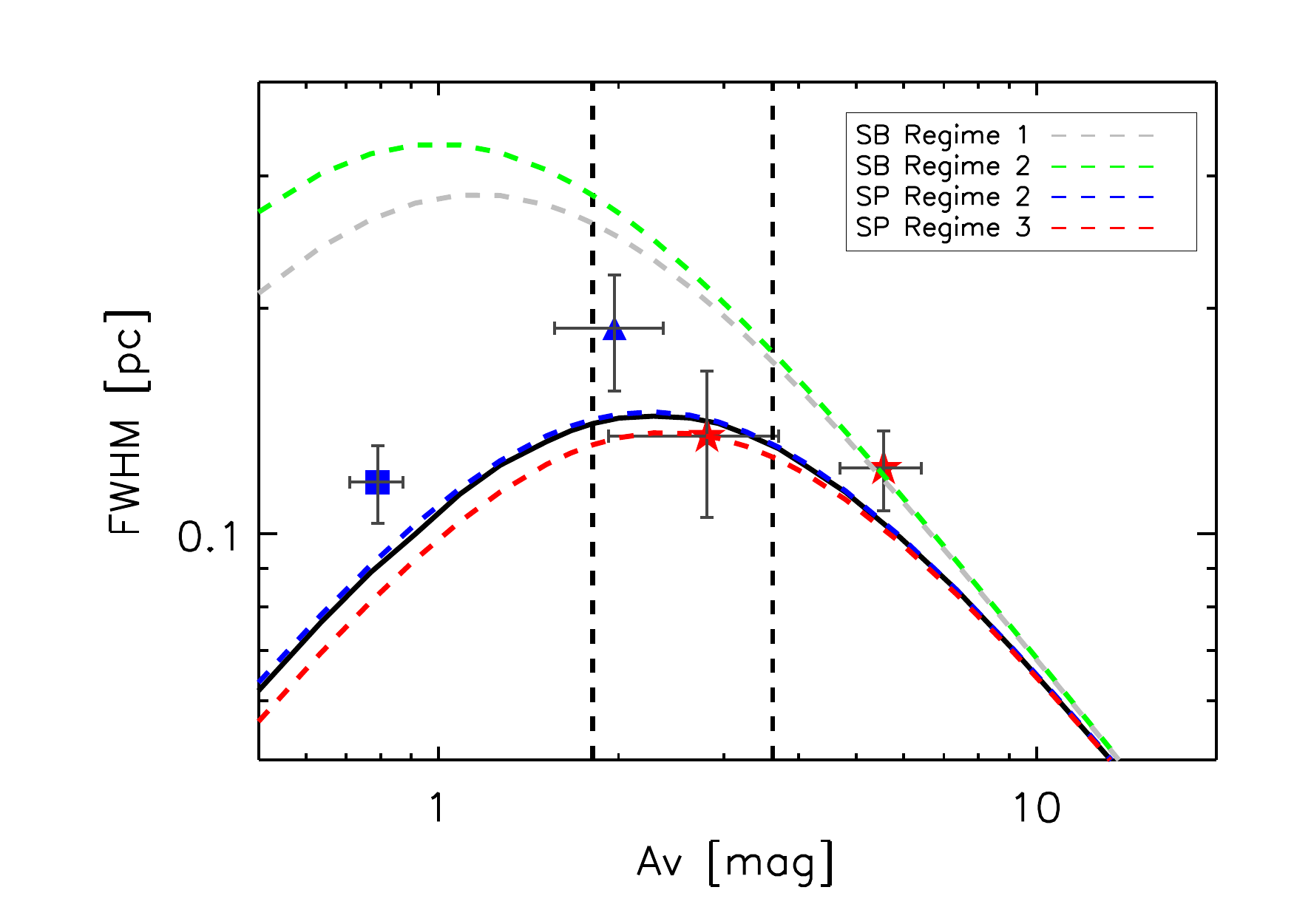}
}\\%
\caption{Panel [a]: Mean filament FWHM as a function of central intrinsic (crest, background-subtracted) filament \av. Black solid curve is the model for isothermal pressure-confined filaments from \citet{fischera2012a} for an external pressure $p_{\mathrm{ext}}$/$k$=2$\times10^4$\,K\,cm$^{-3}$, i.e., environment \av $\approx$2.8\,mag for $\mu=2.33$. Blue and red symbols represent the average change of filament width and crest \av\ for the transition from a \core-dominated subcritical state (SB) to a wing-dominated supercritical one (SP). The average background for each filament group (\av\ (Env.); right axis), derived from the background maps provided by \getsources\ (Section \ref{sec:methods}) is indicated with a black symbol at the characteristic crest \av\ of the group. The clear increase in environment column density with regime (crest \av) is highlighted with a green arrow. Vertical dashed lines mark the approximate boundaries of Regime 2 and Regime 3 for a filament with FWHM$\approx0.13$\,pc. Boundaries would shift to lower \av\ for larger FWHM. Error bars are the standard error on the mean for each type of filament. Panel [b]: Same as Panel [a], but highlighting the theoretical curves corresponding to the average background of each group. Models were derived using the equations included in \citet{fischera2012a}.} 
%}
\label{fig:theory}
\end{figure}

Supercritical (massive) filaments are found in Regime 2 ($4.2\le$\mcore$\le8.4$\,$M_{\odot}$\,pc$^{-1}$) 
and Regime 3 (\mcore$>8.4$\,$M_{\odot}$\,pc$^{-1}$), each 
associated with a different mean environmental column density level and 
filament properties (e.g., Table \ref{table:regimes}; Fig. 9 of Paper I). 
Figure \ref{fig:theory} summarises the mean filament 
width (FWHM) and central ridge column density (here expressed as \av) of 
supercritical filaments in each of the two regimes. 
As a reference, Fig. \ref{fig:theory} also includes the predicted  
FWHM$-$\av\ equilibrium model for self-gravitating, accreting, 
pressure-confined filaments from \citet{fischera2012a}. 
The theoretical curve in Fig. \ref{fig:theory1} corresponds to the default pressure used by these authors in Fig.10 of their work, 
which corresponds to an environment \av\ of $\approx$2.8\,mag as estimated with Eq. A1 of \citet{fischera2012a}. 
Changes on the model curve attributed to variations in external pressure (cloud environment) have been highlighted Fig. \ref{fig:theory2}.

The supercritical population is most dominant in Regime 3, 
which points towards two main requirements characterising the realm of 
supercritical filaments. A filament will grow, collapse, and form protostellar 
objects when it is associated with:
\begin{itemize}
\item a moderately self-gravitating configuration with a 
\core\ component already relatively close to 
supercritical state (e.g.,  a critical value for the filament ridge of \av$\sim5.5$\,mag 
(background-subtracted), or \mcore$\ga$\mcrit$/2$ (Paper I);  in agreement with previous studies, e.g., \citealp{fischera2012a}), and

\item relatively dense environment (\av$\ga3$\,mag; Table \ref{table:regimes}). 
\end{itemize}
Both criteria establish tight observational constraints for modelling the formation of supercritical filaments in simulations.

Interestingly, the two critical column density values (filament and environment) add to a total extinction value similar to the 
proposed threshold for star formation of \av$\sim8$\,mag (e.g., \citealp{andre2010,andre2014}), which further 
supports a transition from subcritical to supercritical 
state linked to a very specific combination of filament and environmental properties. 
The most obvious effect arising from the fulfilment of both criteria would be associated 
enhanced accretion due to the 
significant gravitational potential of the system and the local availability of material. 
The shorter build-up timescales for massive systems in dense environments (e.g., \citealp{kirk2015}) 
would also allow the subcritical-supercritical transition to occur while fragmentation and star 
formation develop ($t\la1$\,Myr). This scenario is consistent with the findings 
from \citet{vanloo2014} and \citet{schisano2014}, who suggest that filaments 
initiate fragmentation while still in the formation stage. 
Considering short fragmentation timescales relative to 
accretion (e.g., \citealp{heitsch2013a}), a subcritical filament with \mcore$\sim8.5$\,$M_{\odot}$\,pc$^{-1}$ 
(maximum \mcore\ for any subcritical filament; Paper I) would quickly fragment and reach 
supercritical level in $\sim10^5-10^6$\,yr for accretion rates of 
$10^{-4}-10^{-5}$\,$M_{\odot}$\,pc$^{-1}$\,yr$^{-1}$ (e.g., \citealp{schisano2014}; \citealp{palmeirim2013}).
Taking into account the estimated lifetimes of prestellar cores 
($\sim10^{6}$\,yr, e.g., \citealp{andre2014}), 
our results are therefore consistent with prestellar substructure forming when filaments 
reach a significantly self-gravitating stage at \mcore\ level. This substructure would 
then evolve as the filament continues assembling its mass beyond supercritical threshold, 
ultimately leading to the presence of accreting supercritical filaments 
associated with protostars and already active star-formation (in agreement with results from e.g., \citealp{toala2012}).

\subsubsection{The path towards supercriticality}
Identification of the precursors of the supercritical filaments and their most probable 
evolutionary sequence depends on the assumed filament lifetime and 
the evolution of the filament with its environment.
In Fig. \ref{fig:theory} we have highlighted the location of subcritical filaments in Regime 1 and Regime 2 
that satisfy the criteria and our established assumptions for being potential precursors of supercritical filaments. 
The position of subcritical filaments in Regime 1 (R1) in the figure corresponds to the mean 
\av$-$FWHM properties of filaments in this regime that have \mcore, \mwing, 
ridge \av, and environment \av\ lower than supercritical wing-dominated filaments in Regimes 2 (R2) and 3 (R3) 
(marked with star symbols in the diagram).  Similarly, the point signalling the location of subcritical filaments in R2
traces the mean \av$-$FWHM properties of  the core-dominated subcritical structures with linear mass 
densities and environmental column density below that of the wing-dominated supercritical filaments in R3.

\begin{itemize}
\item Filaments in Regime 1 are associated predominantly with the most diffuse (\av$\sim1.5$\,mag) environments 
and low central column densities barely at, or below, those 
required for reasonable self-gravitating structures (Table \ref{table:regimes}).  
They are also associated with the narrowest widths, which is consistent 
with the predictions from the (magneto) hydrodynamical theoretical models of \citet{hennebelle2013b}.

\item Subcritical filaments in Regime 2 (environmental \av$\sim2.5$\,mag) approach the turnover point of 
the FWHM-\av\ curve in Fig. \ref{fig:theory}. 
Filaments in these denser environments 
are systematically associated with higher \core\ and wing linear mass densities (Paper I) and 
are therefore reasonably self-gravitating structures.
\end{itemize}

Under our main assumption of filament evolving by accretion (\mtot\ increasing with time; e.g., Fig. \ref{fig:model1}), 
the properties of filaments in the different regimes highlighted in Fig. \ref{fig:theory} are 
strongly suggestive of a filament-environment co-evolutionary scenario. 
Filament growth, inferred by the increase in 
\mcore, \mwing, and ridge \av\ from Regime 1 to Regime 3, 
appears to be intimately associated with an increase in environmental \av, 
although the later changing by a smaller degree:  
 $<$\mcore$>$ and ridge $<$\av$>$ change by a factor of $\sim$6, $<$\mwing$>$ by $\sim$8.5, and 
 the environmental column density increases just by a factor of $\sim$2 (Table \ref{table:regimes}).

\citet{kirk2015} reported a good agreement between `star-forming' filaments (profiles) and the 
models from \citet{fischera2012a}. 
The also apparently good agreement of the pressure-confined equilibrium model shown in Fig. \ref{fig:theory1} 
with the supercritical filaments in this work (particularly in Regime 2) 
can be explained due to the environment \av\ of these filaments being relatively 
similar to that used for that particular model curve (\av$\approx$2.8\,mag, c.f., $<$\av$>$=2.7$-$2.9\,mag for Regimes 2 and 3, respectively). 
We note that such dense environments appear to be typical of clouds with active star formation (e.g., \citealp{rivera2015}). 
Subcritical filaments, on the other hand, are associated with a much lower environment \av, which could explain their 
differences with respect to the equilibrium curve in Fig. \ref{fig:theory1}. 
We observe, however, that their properties still differ from those expected for equilibrium, pressure-confined filaments in their same 
diffuse environments  (Fig. \ref{fig:theory2}; grey and green curves). 
Their widths would be consistent with a much higher external pressure than the one predicted based on their observed environment \av. 
The observed discrepancy can arise due to multiple factors: different environmental temperature and chemical 
composition in diffuse media, inclination effects, or different mechanisms leading to out-of-equilibrium filamentary structures 
in diffuse environments (e.g., stellar feedback, magnetic fields, turbulence). Indeed, the local structure of many filaments in diffuse regions in the GCC sample 
appear to be intrinsically associated with triggered regions, such as swept-up structures and cometary globules. 
Furthermore, the approximation used by \citet{fischera2012a} to derive $p_{\mathrm{ext}}/k$ based on environment \av\ assumes 
relatively dense, self-gravitating, and predominantly molecular clouds resembling pressure-confined isothermal spheres, 
which is an unrealistic approximation for the most diffuse fields in the GCC associated with subcritical filaments.
The location and strong influence of external events on diffuse filaments therefore argue against the presence of 
a dominant population in equilibrium state in low density media, with the non self-gravitating population ultimately dispersing with time 
without undergoing star-formation (Fig. \ref{fig:model2}).

Despite the observed similarities with Regime 2 and, to some extent,  Regime 3, the assumption of an 
equilibrium state for these structures is also questionable. 
First, structures exist with linear mass density above the critical value for equilibrium (Eq. 4; in \citealp{fischera2012a}), 
and filaments resolved by \herschel\ have been frequently observed to be composed of 
bundles of smaller scale velocity-coherent filaments (e.g., \citealp{hacar2013}). 
Second, while equilibrium cases can well exist in non (or slowly) evolving environments 
(e.g., arising from turbulent dissipation in the diffuse ISM; e.g., \citealp{tafalla2015}), 
such stationary models would fail to predict a potential evolution of the filament properties with 
its environment as inferred from the \herschel\ data for the formation of star-forming filaments. 

Numerical evolutionary models in large scale simulations provide a suitable alternative for exploring feasible scenarios consistent with 
observations. Indeed, the simultaneous increase of environmental level with filament growth for the type of structures observed in our data is more 
reminiscent of the dynamically evolving filaments predicted to arise within clouds undergoing large-scale collapse. 
Numerical models that describe the different stages of evolution of such clouds (e.g., \citealp{gomez2014}; \citealp{kirk2015}) 
show that filaments may evolve from a subcritical to a supercritical regime as a function of time. Furthermore, reported column density 
maps and animations seem to support a co-evolution of \mcore\ and environment, although actual quantification of such evolutionary 
process has not been reported, to our knowledge.
Some subcritical filaments in Regime 1, as well as subcritical and 
supercritical filaments in Regime 2, can become precursors of 
Regime 3 supercritical filaments if associated with regions of 
enhanced potential wells, i.e., benefitting from a particularly rapid and/or continuous process of mass assembly. 
In numerical simulations of clouds undergoing large scale collapse, massive filaments 
with central column densities similar to those of Regime 2 and Regime 3 
are present at relatively late evolutionary stages, in the densest regions arising from the converging inflows
($t\sim10-20$\,Myr, e.g., \citealp{vazquez2007}; \citealp{heitsch2008b}; \citealp{gomez2014}). 
 Once the critical filament \mcore\ and environmental conditions are reached, star-formation and evolution 
 progresses quickly due to the gravitational acceleration and significant mass accumulation rate associated 
 with these conditions (e.g., \citealp{heitsch2009}). 
 
 Filaments in Regime 2 (Fig. \ref{fig:theory}) are associated with relatively modest star-forming potential 
 ($<$\mcore$>$$\approx5.5$\,$M_{\odot}$\,pc$^{-1}$; Table \ref{table:regimes}). 
Taking into account the average \mline\ associated by default with `star-forming' filaments (\mcrit), 
Regime 2 supercritical filaments are more consistent with being overall quiescent structures, 
capable of just sporadic and isolated star-forming events. 
However, those filaments located in growing potential wells would further benefit from a more significant, 
continuous and large-scale mass-inflow, leading to a fast additional increase in \mtot\ (therefore \mcore\ and \mwing) and a 
transition to a fully supercritical star-forming state (Regime 3) in short $t\sim1$\,Myr prestellar-protostellar 
transition timescales (e.g., \citealp{heitsch2008b}; \citealp{schisano2014}).
The lack of publicly available models fully describing the simultaneous 
evolution in time of \mcore, \mwing, filament crest \av, width, and environmental \av\ prevents a detailed comparison of these 
observations predictions with simulations. However, the numbers available from accretion-based models 
do support the interpretation of the \herschel\ data described in this work. 

A well defined filament with central crest \av\ comparable to those of subcritical filaments in Regime 2 
is identified in the simulations from \citet{gomez2014} at e.g., X,Y=2,-2 pc and $t=24.4$\,Myr in animation (a), Fig. 3 of that paper. 
Assuming a similar background as those in Regime 2 and similar dust properties in the column density calculation, 
this filament increases its central \av\ by at least a factor of 2 by $t=25.5$\,Myr. The resulting change in 
\av\ is remarkably similar to that observed between supercritical filaments in Regime 2 and Regime 3, before the filament is swept into 
the central regions of the cloud by the large scale collapse of the cloud in the simulations.
Within this same $\sim1$\,Myr timescale, filaments in the simulations from \citet{smith2014}  are also observed to 
increase their linear mass density by a factor $\ga3$, at comparable filament widths (constant or moderately decreasing- Figs. \ref{fig:model1}; \ref{fig:model3}), during their transition from subcritical to supercritical state. 
The trends observed in simulations under large scale collapse are therefore in line with our observationally-based 
predictions of a transition to supercritical state within Regime 2 and extending to Regime 3. 

Our claim of Regime 2 and even Regime 1 subcritical filaments being possible precursors of the most massive Regime 3 supercritical 
filaments is equally in tune with the same numerical models.
Our chosen filament in the simulations from \citet{gomez2014}  evolves in position and location 
with the inwards flow of the cloud collapse, but the filament can be traced during 
several Myr (animation (b)). The `long-lived' nature of these structures and the tendency of the environment to evolve to a 
denser state as a function of time with the aid of gravity \citep{kirk2015} allow for the filament-environment 
co-evolution scenario implied in Fig. \ref{fig:theory}. 
The similarities of the average filament properties with those expected from 
an equilibrium configuration would still be compatible with evolution if such configuration 
is established locally by the pressure balance with the evolving environment, and in timescales much shorter than 
that of large scale collapse of the region. This scenario would also explain the better agreement of equilibrium-based estimates 
with supercritical filaments at moderate densities, and the larger differences with the 
most diffuse (some likely dispersing) and the very dense (fast evolving) structures.

A final piece of evidence in support of the large scale gravitationally driven scenario in the formation of supercritical filaments 
in molecular clouds is found in the filament profile change as a function of time presented in Paper I. 
The tendency of filaments to develop pronounced wing components (\mwing), relative to their filament cores (\mcore), as 
they approach supercritical state is reminiscent of the behaviour shown by collapsing cores within a collapsing cloud 
in the models from \citet{naranjo2015}. A similar behaviour for the `large-scale' (filament) component of the core is likely, 
due to it being an intrinsic part of the filament structure itself.

The observational model inferred by our \herschel\ observations therefore 
not only provides a self-consistent picture in agreement with simulations, but 
it also agrees well with the model of dynamic star formation and the presence 
of a spread in stellar ages in star-forming regions (e.g., \citealp{hartmann2012}; \citealp{zamora2012}).
We note, however, that the fate of the filament will depend on how, where, and when the filament is formed, 
and not all filaments are destined to become supercritical. 
In a relatively constant diffuse environment 
(e.g., off-cloud locations and diffuse ISM, high-galactic latitudes), 
filaments will be quiescent and barely self-gravitating. Evolution and collapse by self-gravity of 
primordially low-mass filaments in these environments can be prevented by stabilising processes 
and by the limited accretion arising from their weak gravitational potential and the low availability of material. 
These structures would therefore be predominantly transient, expanding and dispersing before the onset of star-formation. 
For many of these low-mass filaments, the lack of significant accretion 
would be consistent with the small internal velocity dispersion observed 
in molecular observations \citep{arzoumanian2013}. 
Triggering could instead be 
the key driver leading to star-formation in low column density environments 
where gravity cannot initially play a significant role, 
and where wing development and accretion are therefore severely limited.  

Ultimately, filament formation and evolution is a complex interplay of 
intrinsic and environmental conditions, most likely driven by out-of-equilibrium processes, 
and results from the \herschel\ datasets should be put in context with 
complementary observational studies in order to develop a 
full model of star formation in filaments. 
Molecular line observations will be particularly needed to 
constrain the dynamical state of the filament and its environment. 
Such observations are critical in order to fully validate the proposed co-evolution model.

Massive supercritical filaments  
(and their subcritical progenitors) could also exhibit other exclusive 
properties not traced by our data.
Our results, for instance, do not exclude the possibility of 
magnetic fields playing an important role in filament formation. 
The orientation of magnetic fields has been observed to be linked to 
column density (e.g., \citealp{planckxxxii,planckxxxv}; \citealp{malinen2015}), with 
star-forming filaments having a preferred axis orientation with 
respect to the local magnetic field (e.g., \citealp{andre2014}).
Similarly, the uncertainty level of our width measurements relative to the small 
changes between the different Regimes  
implies that results should be taken with caution 
when trying to identify a characteristic width behaviour for our 
(scarce) supercritical sample. 
A more significant sample of \core-dominated supercritical filaments, 
combined with higher resolution observations and 
numerical simulations, will aid in further constraining the 
physical conditions and timescales associated with the evolution of 
filaments already in supercritical state and actively forming new stars.

\section{Conclusions}\label{sec:conclusion}
A subsample of  \herschel\ fields of the Galactic Cold Cores Programme at $D<500$\,pc 
have been used to investigate the observational signatures associated with 
the onset of star formation in filaments. 

Filaments in different environments were identified and extracted with  
 the \getfilaments\ algorithm. Physical properties (linear mass density, width, crest column density), 
 as well as the structural components of the filament (the `core' innermost region 
 and the `wing' power-law like component at large radii of the filament profile) were 
 quantified as described in Paper I. The characteristics derived from the \nh\ profile 
 fitted with a Plummer-like function were investigated as a function of environment 
 and stability (\mtot). The analysis was performed in order to constrain an 
 observationally-based evolutionary model 
 from a subcritical to a supercritical (unstable) state that can be put in context 
 with theoretical models of filament formation and evolution.
 
The combination of local environment and \mcore\ at a given time has 
a critical role in determining the evolutionary path 
and ultimate fate (star-forming potential) of a given filament in $t\la1$\,Myr. 
Only strongly self-gravitating structures with a relative massive \core\ component associated 
with a dense environment (\av$\ga3$\,mag) have the potential for becoming supercritical and star-forming in timescales 
comparable to the observed lifetime of prestellar sources. 
Filament evolution via mass assembly progresses with little changes in filament average width 
($<$FWHM$>\sim0.13$\,pc). Low-mass filaments in diffuse environments tend to increase their 
FWHM with time, while self-gravitating filaments in denser regions contract or evolve at similar width 
during the transition to supercriticality. 

Our proposed observationally-based model for the formation of supercritical star-forming filaments 
has been compared with predictions from numerical models of filament evolution. 
While filamentary properties for filaments in relatively dense media 
are similar to those predicted for pressure-confined filaments in equilibrium, 
the non-negligible differences with respect to the \herschel\ measurements and the 
simultaneous growth of the filament linear mass density with its environment with time implied by the \herschel\ 
data are more consistent with dynamical models of cloud evolution. 
The filament$-$environment coevolution scenario is an intrinsic feature of filaments in clouds undergoing 
large scale collapse. Those structures formed in the potential wells that emerge during the convergent inflow motions 
benefit from enhanced accretion and efficient mass assembly. The physical changes of filaments during the 
subcritical-supercritical transition predicted by simulations are quantitatively consistent with 
those derived from observations.

While the average properties of a filament population can reveal clues as to their formation 
and evolution, the dispersion of filament properties observed in the GCC fields remains significant. 
Similarly, the feasibility of the evolutionary paths considered are highly dependent on the established assumptions, 
and the evolution itself can also be highly dependent on external conditions, such as Galactic position 
and physical processes forming the filament.
A complete understanding of filament structure and 
evolution must not only investigate the impact of processes such as 
shocks, turbulence, and magnetic fields, together with gravity, but the observed properties must also 
be put in context with local environment, star-forming activity, and history of the region. 
In-depth studies of these properties, even for individual fields, will be key 
for understanding the wide diversity of the filament population.

%%%%%%%%%%%%%%%%%%%%%%%%%%%%%%%%%%%%%%%%%%%%%%%%%%%%%%%%%%%%%%%%%%%%%%%%%%%%
\begin{acknowledgements}
A.R-I. is currently a Research Fellow at ESA/ESAC and acknowledges support from the ESA Internal Research Fellowship Programme. 
The authors would like to thank Enrique V\'{a}zquez-Semadeni for his in-depth study of our results and detailed discussions which have greatly improved the content and presentation of this work. We are also grateful to Joerg Fischera for providing valuable insight for the interpretation of filament models. We thank PCMI for its general support to the `Galactic Cold Cores' project activities.
J.M. and V.-M.P. acknowledge the support of Academy of Finland grant 250741. 
M.J. acknowledges the support of Academy of
Finland grants 250741 and 1285769, as well as the Observatoire Midi-Pyrenees (OMP) in Toulouse for 
its support for a 2 months stay at IRAP in the frame of the `OMP visitor programme 2014'. 
L.V.T. acknowledges OTKA grants NN111016 and K101393. 
We also thank J. Fischera, D. Arzoumanian, E. Falgarone, and P. Andr\'{e} for useful discussions.
SPIRE has been developed by a consortium of institutes led by Cardiff Univ. (UK) and including: Univ. Lethbridge (Canada); NAOC (China); CEA, LAM (France); IFSI, Univ. Padua (Italy); IAC (Spain); Stockholm Observatory (Sweden); Imperial College London, RAL, UCL-MSSL, UKATC, Univ. Sussex (UK); and Caltech, JPL, NHSC, Univ. Colorado (USA). This development has been supported by national funding agencies: CSA (Canada); NAOC (China); CEA, CNES, CNRS (France); ASI (Italy); MCINN (Spain); SNSB (Sweden); STFC, UKSA (UK); and NASA (USA). 
PACS has been developed by a consortium of institutes led by MPE (Germany) and including UVIE (Austria); KU Leuven, CSL, IMEC (Belgium); CEA, LAM (France); MPIA (Germany); INAF-IFSI/OAA/OAP/OAT, LENS, SISSA (Italy); IAC (Spain). This development has been supported by the funding agencies BMVIT (Austria), ESA-PRODEX (Belgium), CEA/CNES (France), DLR (Germany), ASI/INAF (Italy), and CICYT/MCYT (Spain).
\end{acknowledgements}

%%%%%%%%%%%%%%%%%%%%%%%%%%%%%%%%%%%%%%%%%%%%%

%%%%%%%%%%%%%%%%%%%%%%%%%%%%%%%

\bibliographystyle{aa}

%%%%%%%%%%%%%%%%%%%%%%%%%%%%%%%%%%%%%%%%%%%%%%%%%%%%%%%%%%%%%%%%%%%%%%%%%%%%

\end{document}